\documentclass[english]{article}
\usepackage[T1]{fontenc}
\usepackage[latin9]{inputenc}
\usepackage{amssymb}

\usepackage{amsmath}
\usepackage{esint}

\makeatletter
\newcommand{\lyxaddress}[1]{
	\par {\raggedright #1
	\vspace{1.4em}
	\noindent\par}
}

\makeatother

\usepackage{babel}
\begin{document}
\title{Hamiltonian Point of View of Quantum Perturbation Theory}
\author{A.D Berm\'udez Manjarres$^{1}$}
\maketitle

\lyxaddress{$^{1}$Universidad Distrital Francisco Jose de Caldas, Cra 7 No.
40B-53, Bogot\'a, Colombia. Email: ad.bermudez168@uniandes.edu.co}
\begin{abstract}
We explore the relation of Van Vleck-Primas perturbation theory of
quantum mechanics with the Lie-series-based perturbation theory of
Hamiltonian systems in classical mechanics. In contrast to previous
works on the relation of quantum and classical perturbation theories,
our approach is not based on the conceptual similarities between the
two methods. Instead, we show that for quantum systems with a finite-dimensional Hilbert space, the Van Vleck-Primas procedure can be recast exactly
into a classical perturbation problem. As a non-obvious consequence, this approach gives a new way of calculating the geometric phase of quantum systems using tools from the theory of classical canonical transformations.
\end{abstract}

\section{Introduction}

Concepts from classical mechanics played an essential role in the
early days of quantum theory. For example, in the influential paper
of Born, Heisenberg, and Jordan \cite{Born}, an adaptation of the concept of canonical transformation is used to discuss approximations
of a given problem using a perturbation procedure. In Heisenberg's words, ''The analogy with the classical Hamilton\textendash Jacobi
technique was the beginning of all efforts; what we did first was just to try to imitate the old methods as closely as we could'' \cite{heisenberg}. 

Nowadays, the formalism of quantum mechanics is well established,
and plenty of research is done without considering classical ideas.
However, there are still cases where developments in quantum mechanics
originate in adaptations of concepts and techniques from classical
mechanics. In the context of perturbation theory, we can find: averaging
theory to avoid secular terms \cite{Q-C0,Q-C1}, normal forms \cite{Q-Cn,Q-C1-1},
analogs to canonical perturbation theory and the KAM theorem \cite{Q-C2},
and time-dependent methods \cite{Q-Cs,Q-Cs2,casas,casas-1}. On the other hand, results from quantum mechanics have also inspired investigations in classical perturbation theory, as the Kato expansion applied to the
Liouville operator \cite{C-Q0,C-Q1} and the use of Feynman diagrams
in classical mechanics \cite{C-Q2,C-Q3}.

\[
\]

This work focuses on the relationship between quantum and classical time-independent perturbation
theories. The central question we address in this work can be stated as follows: let $\hat{H}_{0}$
be a Hamiltonian operator with discrete and non-degenerate spectra.
Assume the eigenvalues $\left\{ E_{n}\right\} $ and eigenvectors
$\left\{ \left|n\right\rangle \right\} $ of $\hat{H}_{0}$ are known.
If $\hat{V}$ is an small perturbation, what are the eigenvalues $\left\{ E'_{n}\right\} $
and eigenvectors $\left\{ \left|n'\right\rangle \right\} $ of $\hat{H}=\hat{H}_{0}+\hat{V}$?
There are several approaches to deal with the quantum perturbation
problem, like Rayleigh-Schr\"odinger \cite{QPT1} or Brillouin-Wigner
\cite{QPT2,QPT3}. We will only consider here the perturbation procedure
known as the Van Vleck-Primas theory (VPP) \cite{vlan vleck 1,van vleck 2,van vleck 3,van vleck 4}\footnote{While Van Vleck-Primas perturbation theory can be used in more general
cases, in this work we only consider the restricted case of Hamiltonian
operators with non-degenerate spectrum. }. 

The equivalent perturbation problem of classical mechanics is the
following: Let $H_{0}$ be a Hamiltonian function with a set of action-angle
variables $(I,\theta)$ such that $H_{0}=H_{0}(I)$. If $V(I,\theta)$ is a small perturbative function, what are the action-angle variables
$(I',\theta')$ of the Hamiltonian function $H=H_{0}(I)+V(I,\theta)$
such that $H=H(I')$? Mainly developed to deal with problems in celestial mechanics, the main methods in classical perturbation theory are the
Von Zeipel-Poincare \cite{arnold,arnold 2}, and the ones based on
Lie series like Hori's \cite{classical0} and Lie-Deprit \cite{calssical,classical2}
(see \cite{classical3} for a complete exposition of the aforementioned methods). While all the classical perturbation theories mentioned give the same result in all orders, they differ in their approach
and usefulness for a given problem. 

Here, we investigate the relationship between VVP and the classical
perturbation theory based on Lie series. The formal similarities between
the quantum and classical procedures are well known, see for example
\cite{Q-C1,casas}. Instead of comparing the two theories as analogs,
we will show that the VVP procedure can be recast exactly into a classical
perturbation theory. We will show that a direct translation of VVP
gives Hori's classical procedure.

The bridge between quantum and classical perturbation theory will
be the Hamiltonian version of quantum mechanics \cite{strocci,heslot}.
The idea is to put quantum dynamics in the same mathematical language
as classical Hamiltonian mechanics. This allows using classical techniques in quantum scenarios \cite{ham,ham2,ham3,ham4}.

Using the Hamiltonian version of quantum mechanics, we will show that, in a quantum system with a finite Hilbert space, the quantum perturbation problem can be restated and solved as a problem in classical perturbation theory.

This work is organized as follows: In section 2, we present the necessary
aspects of the Hamiltonian version of quantum mechanics. We explain
the relation between the Scrh\"odinger equation and the Hamilton equations,
the commutator and the Poisson bracket, and unitary and canonical
transformations. Section 3 deals with the Van Vleck-Primas perturbation
theory. Our presentation of the theory will follow the one found in
\cite{van vleck 3}. The purpose of the section is to present the
theory in a way that will allow a quick recast into the classical
Lie series formalism. In section 4, we will rewrite the quantum perturbation
problem in the mathematical language of classical mechanics. Finally, in section 5, we show the relationship between the generators of Lie transformations and the geometric phase of a quantum system.  We calculate the Berry phase of a spin 1/2 interacting with a magnetic field using classical techniques. 

We set $\hbar=1$ throughout this paper 

\section{Hamiltonian Version of Quantum Mechanics}

In this section, our presentation follows \cite{strocci,heslot}, we refer to those works for the proof of the statements presented here.

Consider a quantum system with a Hamiltonian $\hat{H}$ with a finite spectrum. The evolution of an arbitrary vector $\left|\psi\right\rangle $
is given by the Schr\"odinger equation 

\begin{equation}
i\frac{d}{dt}\left|\psi\right\rangle =\hat{H}\left|\psi\right\rangle .\label{scrho}
\end{equation}
We can write (\ref{scrho}) in classical Hamiltonian form as follows:
Let $\left\{ \left|\phi_{k}\right\rangle \right\} $ be a basis of
orthonormal vectors of the Hilbert space. The vector $\left|\psi\right\rangle $
can be expanded as

\begin{equation}
\left|\psi\right\rangle =\sum_{k}\lambda_{k}\left|\phi_{k}\right\rangle ,\label{expansion}
\end{equation}
where the expansion coefficients $\lambda_{k}$ are complex numbers.
The coefficients can be written in terms of their real and imaginary parts as

\begin{equation}
\lambda_{k}=\frac{q_{k}+ip_{k}}{\sqrt{2}}.\label{coef}
\end{equation}

Inserting Eq.(\ref{expansion}) into Eq.(\ref{scrho}) and taking
the scalar product with $\left\langle \psi\right|$ gives the following
set of Hamiltonian equations for the real and imaginary part of $\lambda_{k}$

\begin{eqnarray}
\frac{dq_{k}}{dt} & = & \frac{\partial H}{\partial p_{k}},\nonumber \\
\frac{dp_{k}}{dt} & = & -\frac{\partial H}{\partial q_{k}},\label{hamilton}
\end{eqnarray}
where the (real-valued) Hamiltonian function is given by 

\begin{equation}
H(q,p)=\left\langle \psi\right|\hat{H}\left|\psi\right\rangle .\label{clahamiltonian}
\end{equation}

The coordinates $(q,p)$ define a phase space $\mathcal{P}$ of twice
the dimension of the Hilbert space. $\mathcal{P}$ is endowed with
a natural bracket operation defined by

\begin{equation}
\left\{ f,g\right\} =\sum_{k}\left(\frac{\partial f}{\partial q_{k}}\frac{\partial g}{\partial p_{k}}-\frac{\partial f}{\partial p_{k}}\frac{\partial g}{\partial q_{k}}\right),\label{poissonbracket}
\end{equation}
where $f$ and $g$ are real-valued functions of $(q,p)$. It can
be shown that Eq.(\ref{poissonbracket}) has all the defining properties
of the Poisson bracket of classical Hamiltonian mechanics. Clearly,
the coordinates $(q,p)$ obey 
\begin{eqnarray}
\left\{ q_{k},p_{l}\right\}  & = & \delta_{kl},\nonumber \\
\left\{ q_{k},q_{l}\right\}  & = & \left\{ p_{k},p_{l}\right\} =0.\label{poisson}
\end{eqnarray}

Just as it was done for the Hamiltonian in (\ref{clahamiltonian}),
we can define a real-valued function $g=g\left(q,p\right)$ for any
self-adjoint operators $\hat{g}$ by

\begin{equation}
g=\left\langle \psi\right|\hat{g}\left|\psi\right\rangle =\sum_{k,l}\frac{g_{kl}}{2}\left[\left(q_{k}q_{l}+p_{k}p_{l}\right)+i\left(q_{k}p_{l}-p_{k}q_{l}\right)\right],\label{g}
\end{equation}
where $g_{kl}=\left\langle \phi_{k}\right|\hat{g}\left|\phi_{l}\right\rangle $.
An important result links the commutator of operators with the Poisson
brackets. Namely, the following holds for any two arbitrary self-adjoint
operators $\hat{f}$ and $\hat{g}$:

\begin{equation}
\left\{ f,g\right\} =-i\left\langle \psi\right|[\hat{f},\hat{g}]\left|\psi\right\rangle .\label{bracketP}
\end{equation}
Eq.(\ref{bracketP}) will be used repeatedly in section 4 to link
VVP perturbation theory with the classical Lie series method.

There exist transformations on $\mathcal{P}$ that are canonical in the sense that they let the bracket $\left\{ f,g\right\} $ invariant.
Among all canonical transformations on $\mathcal{P}$, the subset induced by a unitary transformation on the Hilbert space
is of special interest. Let $U$ be a unitary transformation with an associated unitary operator $\hat{U}$. Let us write the action
of $\hat{U}$ on the basis vectors $\left\{ \left|\phi_{k}\right\rangle \right\} $
as

\begin{equation}
\hat{U}\left|\phi_{k}\right\rangle =\left|\phi'_{k}\right\rangle .\label{Uphi}
\end{equation}
The vector $\left|\psi\right\rangle $ can be expanded in terms of
the new basis $\left\{ \left|\phi'_{k}\right\rangle \right\} $ as 

\begin{eqnarray}
\left|\psi\right\rangle  & = & \sum_{k}\frac{1}{\sqrt{2}}\left(q'_{k}+ip'_{k}\right)\left|\phi'_{k}\right\rangle ,
\end{eqnarray}
hence, $U$ defines a change of coordinates in $\mathcal{P}$

\begin{equation}
U:\,(q,p)\longrightarrow(q',p').\label{Uqp}
\end{equation}
The transformation given by (\ref{Uqp}) is canonical. It has the
property that

\begin{equation}
U\left(\left\{ f,g\right\} \right)=\left\{ U(f),U(g)\right\} ,\label{Ufg}
\end{equation}
or writing (\ref{Ufg}) in terms of the coordinates

\begin{eqnarray}
\left\{ f,g\right\} _{(q,p)} & = & \left\{ f,g\right\} _{(q',p')}.\label{fgpq}
\end{eqnarray}

The VVP perturbation theory deals with unitary transformations $U$ close to the identity. Considering only the first-order expansion
of $\hat{U}$, we can write

\begin{equation}
\hat{U}=1+i\epsilon\hat{W},
\end{equation}
where $\hat{W}$ is an Hermitean operator. The action of $U$ on a
given self adjoint operator $\hat{g}$ is 

\begin{equation}
U:\,\hat{g}\longrightarrow\hat{U}\hat{g}\hat{U}^{\dagger}=\hat{g}'=\hat{g}+i\varepsilon\left[\hat{W},\hat{g}\right].\label{ugw}
\end{equation}
The canonical transformation induced by (\ref{ugw}) on the function
$g$ is 

\begin{equation}
U:\,g\longrightarrow g'=g+\varepsilon\left\{ g,W\right\} ,
\end{equation}
where the generator function of the canonical transformation is $W=\left\langle \psi\right|\hat{W}\left|\psi\right\rangle $.

Finally, we are interested in the basis of eigenvectors of the Hamiltonian
$\left\{ \left|n\right\rangle \right\} $,

\begin{equation}
\hat{H}\left|n\right\rangle =E_{n}\left|n\right\rangle .
\end{equation}
We can write $\left|\psi\right\rangle $ in terms of $\left\{ \left|n\right\rangle \right\} $
as

\begin{eqnarray}
\left|\psi\right\rangle  & = & \sum_{n}\frac{1}{\sqrt{2}}\left(q_{n}+ip_{n}\right)\left|n\right\rangle \nonumber \\
 & = & \sum_{n}\sqrt{I_{n}}e^{-i\theta_{n}}\left|n\right\rangle ,\label{psin}
\end{eqnarray}
where the variables $(\theta,I)$ are defined by

\begin{eqnarray}
q_{n} & = & \sqrt{2I_{n}}\cos\theta_{n},\nonumber \\
p_{n} & = & -\sqrt{2I_{n}}\sin\theta_{n}.\label{pqI}
\end{eqnarray}
In terms of variables (\ref{pqI}), the Hamiltonian function reads

\begin{eqnarray}
H & = & \sum_{n}E_{n}\frac{q_{n}^{2}+p_{n}^{2}}{2}\nonumber \\
 & = & \sum_{n}E_{n}I_{n}.\label{HI}
\end{eqnarray}
The coordinates $(\theta,I)$ are a set of action-angle variables
for $H$. They are canonical variables

\begin{eqnarray}
\left\{ \theta_{k},I_{l}\right\}  & = & \delta_{kl},\nonumber \\
\left\{ \theta_{k},\theta_{l}\right\}  & = & \left\{ I_{k},I_{l}\right\} =0,
\end{eqnarray}
and they obey the Hamilton equations

\begin{eqnarray}
\frac{d\theta_{k}}{dt} & = & \frac{\partial H}{\partial I_{k}}=E_{k},\nonumber \\
\frac{dI_{k}}{dt} & = & -\frac{\partial H}{\partial I_{k}}=0.
\end{eqnarray}
In terms of the action-angle variables, the expectation value of any
operator $\hat{A}$ can be written as

\begin{equation}
A(\theta,I)=\sum_{n,n'}\sqrt{I_{n}I_{n'}}A_{n'n}e^{i(\theta_{n'}-\theta_{n})}\label{matrix}
\end{equation}
where the matrix elements of $\hat{A}$ are given by $A_{n'n}=\left\langle n'\right|\hat{A}\left|n\right\rangle $.

These are all the results we need from the Hamiltonian version of
quantum mechanics. We now proceed to introduce the VVP perturbation
theory.

\section{Van Vleck-Primas perturbation theory}

Following \cite{van vleck 3,van vleck 4}, let us consider a quantum system described by an N-dimensional Hilbert space and a Hamiltonian operator given by

\begin{equation}
\hat{H}=\hat{H}_{0}+\sum_{n=1}^{\infty}\varepsilon^{n}\hat{V}_{n},\label{Hgorro}
\end{equation}
where $\varepsilon$ is a small parameter. We assume that we know
exactly the set of eigenvalues $\left\{ E_{n}^{0}\right\} $ and eigenvectors
$\left\{ \left|n_{0}\right\rangle \right\} $ of $\hat{H}_{0}$

\begin{equation}
\hat{H}_{0}\left|n_{0}\right\rangle =E_{n}^{0}\left|n_{0}\right\rangle .
\end{equation}
We also assume the spectrum of $\hat{H}_{0}$ and $\hat{H}$ to be
non-degenerated for all the values of interest of $\varepsilon$.

The Van Vleck-Primas method consists of looking for a unitary operator

\begin{equation}
\hat{U}=\exp\left(i\hat{W}\right),\label{UW}
\end{equation}
 such that
\begin{eqnarray}
\hat{\widetilde{H}} & = & \hat{U}\hat{H}\hat{U}^{\dagger}=\hat{H}_{0}+\hat{K},\label{H'}\\
\left[\hat{H}_{0},\hat{K}\right] & = & \left[\hat{\widetilde{H}},\hat{K}\right]=0.\label{kh0}
\end{eqnarray}
The operator $\hat{K}$ is called the shift operator because, as we
will see, it has the effect of shifting the energy levels. If the
unitary operator $\hat{U}$ exists and can be found, then the eigenvalues
and eigenvectors of $\hat{H}$ are given by

\begin{eqnarray}
E_{n} & = & E_{n}^{0}+\left\langle n\right|\hat{K}\left|n\right\rangle ,\label{EE0k}\\
\left|n\right\rangle  & = & \hat{U}\left|n_{0}\right\rangle .\label{NUn}
\end{eqnarray}

We will show a formal way to obtain $\hat{U}$ to all orders in $\varepsilon$. For now, we are not worrying about the convergence of the procedure. Let us start by noting that Eq.(\ref{H'}) can be
expanded using the Baker-Campbell-Haussdorf lemma as follows

\begin{eqnarray}
\hat{\widetilde{H}} & = & \hat{U}\hat{H}\hat{U}^{\dagger}=\hat{H}_{0}+i\left[\hat{W},\hat{H}\right]-\frac{1}{2!}\left[\hat{W},\left[\hat{W},\hat{H}\right]\right]\nonumber \\
 &  & -\frac{i}{3!}\left[\hat{W},\left[\hat{W},\left[\hat{W},\hat{H}\right]\right]\right]\ldots\nonumber \\
 & = & \sum_{n=0}^{\infty}\frac{i^{n}}{n!}D_{\hat{W}}^{n}\left(\hat{H}\right),\label{DWH}
\end{eqnarray}
where the superoperator $D_{\hat{W}}$ is defined by

\begin{equation}
D_{\hat{W}}=\left[\hat{W},\cdot\right].
\end{equation}
Moreover, we assume we can expand $\hat{K}$ and $\hat{W}$ in a power
series as

\begin{eqnarray}
\hat{W} & = & \varepsilon\hat{W}_{1}+\varepsilon^{2}\hat{W}_{2}+\varepsilon^{3}\hat{W}_{2}+\ldots,\nonumber \\
\hat{K} & = & \varepsilon\hat{K}_{1}+\varepsilon^{2}\hat{K}_{2}+\varepsilon^{3}\hat{K}_{2}+\ldots.\label{power}
\end{eqnarray}
Inserting Eqs.(\ref{DWH}) and (\ref{power}) in Eq. (\ref{H'}), the following equation is obtained

\begin{eqnarray}
\hat{H}_{0}+\sum_{n=1}^{\infty}\varepsilon^{n}\hat{K}_{n} & = & \sum_{k=0}^{\infty}\frac{i^{n}}{n!}D_{\hat{W}}^{n}\left(\hat{H}_{0}+\sum_{n=1}^{\infty}\varepsilon^{n}\hat{V}_{n}\right).\label{hkd}
\end{eqnarray}
Equating terms of the same order in $\varepsilon$ in Eq. (\ref{hkd}), it is possible to write the following equations

\begin{equation}
i\left[\hat{H}_{0},\hat{W}_{n}\right]=\hat{\Psi}_{n}-\hat{K}_{n},\label{Qhomological}
\end{equation}
where the first three $\hat{\Psi}$ are given by

\begin{eqnarray}
\hat{\Psi}_{1} & = & \hat{V_{1}}\nonumber \\
\hat{\Psi}_{2} & = & \hat{V_{2}}-\frac{1}{2}\left[\hat{W}_{1},\left[\hat{W}_{1},\hat{H}_{0}\right]\right]+i\left[\hat{W}_{1},\hat{V_{1}}\right]\nonumber \\
\hat{\Psi}_{3} & = & \hat{V_{3}}-\frac{1}{2}\left[\hat{W}_{2},\left[\hat{W}_{1},\hat{H}_{0}\right]\right]-\frac{1}{2}\left[\hat{W}_{1},\left[\hat{W}_{2},\hat{H}_{0}\right]\right]\nonumber \\
 &  & -\frac{i}{6}\left[\hat{W}_{1},\left[\hat{W}_{1},\left[\hat{W}_{1},\hat{H}_{0}\right]\right]\right]+i\left[\hat{W}_{2},\hat{V_{1}}\right]\nonumber \\
 &  & -\frac{1}{2}\left[\hat{W}_{1},\left[\hat{W}_{1},\hat{V_{1}}\right]\right]+i\left[\hat{W}_{1},\hat{V_{2}}\right].\label{psi123}
\end{eqnarray}
The Eqs. (\ref{Qhomological}) can be simplified further. Let the parallel-projection
linear superoperator $\pi$ be defined by giving its action on an
arbitrary $\hat{X}$ as

\begin{equation}
\pi(\hat{X})=\sum_{n=1}^{N}\left\langle n_{0}\right|\hat{X}\left|n_{0}\right\rangle \left|n_{0}\right\rangle \left\langle n_{0}\right|.\label{defdiag}
\end{equation}
Any arbitrary operator $\hat{X}$ will commute with $\hat{H}_{0}$
if and only if $\pi(\hat{X})=\hat{X}$. Moreover, the following identity
can be shown to always hold for any $\hat{X}$,

\begin{equation}
\pi\left(\left[\hat{H}_{0},\hat{X}\right]\right)=0.\label{picom}
\end{equation}
Since the shift operator is defined by Eq.(\ref{kh0}) to commute
with $\hat{H}_{0}$ , we can use the linearity of $\pi$ and the identity
(\ref{picom}) to take the parallel-projection of Eq.(\ref{Qhomological})
and obtain the following

\begin{equation}
\pi\left(\hat{K}_{n}\right)=\hat{K}_{n}=\pi\left(\hat{\Psi}_{n}\right).\label{piK}
\end{equation}
Hence, Eq (\ref{piK}) reduces to

\begin{equation}
i\left[\hat{H}_{0},\hat{W}_{n}\right]=\hat{\Psi}_{n}-\pi\left(\hat{\Psi}_{n}\right).\label{Qhomological-1}
\end{equation}
Notice that it is necessary to solve Eqs. (\ref{Qhomological-1}) in ascending order. To find $\hat{W}_{n}$, all the $\hat{W}_{i}$ from
$\hat{W}_{1}$ to $\hat{W}_{n-1}$ have first to be known.

To end this section, we point out that the operation $\pi\left(\hat{\Psi}_{n}\right)$
can be written as a time average in the following way \cite{jauslin}.

\begin{eqnarray}
\pi\left(\hat{\Psi}_{n}\right) & = & \lim_{\tau\rightarrow\infty}\frac{1}{\tau}\int_{0}^{\infty}dt\,e^{-i\hat{H}_{0}t}\hat{\Psi}_{n}e^{i\hat{H}_{0}t}.\label{qaverage}\\
\hat{W}_{n} & = & \lim_{\tau\rightarrow\infty}\frac{1}{\tau}\int_{0}^{\infty}dt\,\int_{0}^{t}ds\,e^{-i\hat{H}_{0}s}\left(\hat{\Psi}_{n}-\pi\left(\hat{\Psi}_{n}\right)\right)e^{i\hat{H}_{0}s}\nonumber \\
 & = & -i\sum_{k\neq k'}\frac{\left\langle k_{0}\right|\hat{\Psi}_{n}\left|k'_{0}\right\rangle }{E_{k}^{0}-E_{k'}^{0}}\left|k'_{0}\right\rangle \left\langle k_{0}\right|.\label{psinint}
\end{eqnarray}

\section{Quantum perturbation as a classical theory}

We define a ``classical'' function of the Hamiltonian operator (\ref{Hgorro})
by taking its expectation value according to the rule given by Eq.(\ref{g}).
Using $\left|\psi\right\rangle =\sum_{n}\sqrt{I_{n}^{(0)}}e^{-i\theta_{n}^{(0)}}\left|n_{0}\right\rangle $,
we can evaluate the expectation value $H=\left\langle \psi\right|\hat{H}\left|\psi\right\rangle $as 

\begin{eqnarray}
H(I,\theta) & = & H_{0}+\sum_{n=1}^{\infty}\varepsilon^{n}\left\langle \psi\right|\hat{V}_{n}\left|\psi\right\rangle \nonumber \\
 & = & \sum_{k=1}^{N}E_{n}^{0}\left(\frac{(q_{k}^{0})^{2}+(p_{k}^{0})^{2}}{2}\right)+\sum_{n=1}^{\infty}\varepsilon^{n}V_{n}(q^{0},p^{0})\nonumber \\
 & = & \sum_{k=1}^{N}E_{n}^{0}I_{n}^{(0)}+\sum_{n=1}^{\infty}\varepsilon^{n}V_{n}(I^{(0)},\theta^{(0)}).\label{classicalh}
\end{eqnarray}
A classical system described by a Hamiltonian of the form of Eq.(\ref{classicalh}) is said to be quasiharmonic. The energies $E_{n}^{0}$ play the role
of frequencies of the unperturbed oscillators. There is no problem
if the energies (frequencies) are negative numbers. However, in what follows, we cannot allow any $E_{n}^{0}$ to vanish. This does not
present a problem since, without changing the quantum dynamics, we
can always redefine $\hat{H_{0}}$ by adding a constant multiple of the identity so that no energy equals zero.

The goal of classical Hamiltonian perturbation theory is to find a
time-independent canonical transformation into a new set of action-angle
variables

\begin{equation}
T:(\theta^{(0)},I^{(0)})\rightarrow\left(\theta,I\right),\label{T}
\end{equation}
such that the perturbed Hamiltonian becomes a function only of the new actions\footnote{Here we use the conservation of the Hamiltonian under time-independent
canonical transformation.}

\begin{equation}
H^{*}\left(I\right)=H\left(\theta^{0},I^{0}\right).\label{H*}
\end{equation}
Here and hereafter, the $^{*}$ will remind us that the function involved
have to be considered as a function of the new action-angle variables
i.e., $f^{*}\left(\theta,I\right)=f(\theta^{(0)}\left(\theta,I\right),I^{(0)}\left(\theta,I\right))$.
Now, if such a transformation can be found, the Hamilton equations of
motion becomes

\begin{eqnarray}
\frac{d\theta_{i}}{dt} & = & \frac{\partial H^{*}}{\partial I_{i}}=E_{i},\nonumber \\
\frac{dI_{i}}{dt} & = & -\frac{\partial H^{*}}{\partial\theta_{i}}=0.
\end{eqnarray}

Staying within the classical formalism, the canonical transformation
that allows a solution\footnote{By this we mean a formal solution. Classical perturbation theory
is famous for its problem with small denominators that can cause terms
in the perturbation expansion to diverge.} of the perturbed Hamiltonian can be investigated using Lie series.
However, we already know the unitary transformation that solves the
quantum problem, at least formally. The associated canonical transformation
of the unitary operator (\ref{UW}) is the solution to the classical
problem (\ref{H*}), as we shall see. The vector $\left|\psi\right\rangle $
can be written in the basis given by Eq.(\ref{NUn}) as $\left|\psi\right\rangle =\sum_{n}\sqrt{I_{n}}e^{-i\theta_{n}}\left|n\right\rangle =\sum_{n}\sqrt{I_{n}}e^{-i\theta_{n}}(\hat{U}\left|n_{0}\right\rangle )$.
Hence, Taking the expectation value of $\hat{H}$ in this basis would
give 
\begin{equation}
H^{*}=\sum_{n=1}^{N}E_{n}\left(\frac{q_{k}^{2}+p_{k}^{2}}{2}\right)=\sum_{n=1}^{N}E_{n}I_{n},\label{hnewbasis}
\end{equation}
where the perturbed energies $E_{n}$ are given by Eq.(\ref{EE0k}).
The Eq.(\ref{hnewbasis}) correspond to the desired Hamiltonian function
(\ref{H*}). The Eq.(\ref{hnewbasis}) is said to be the Birkhoff
normal form of $H^{*}$. As $\hat{H}$ should always be diagonalizable,
there should always exist variables $\left(\theta,I\right)$ such
that $H^{*}$ can be put in normal form. This is, $\hat{H}$ is always
integrable in the Liouville-Arnold sense.

The concept of resonance in the frequencies is of fundamental importance in classical perturbation theories. The non-degeneracy of the quantum
Hamiltonian means that $E_{n}^{0}\neq E_{m}^{0}$. In the classical jargon, it is said that there is no \emph{trivial} degeneracy in the
energies (frequencies). The energies are called to be non-resonant if for all integers $k_{n}$ (not all equal to zero) we have that
\begin{equation}
\mathbf{k}\cdot\mathbf{E}^{0}=\sum_{n=1}^{N}k_{n}E_{n}^{0}\neq0.\label{nonresonance}
\end{equation}
On the other hand, a resonance is called of order $l$ when there
exist a vector of integers $\mathbf{k}$ such that

\begin{eqnarray*}
\mathbf{k}\cdot\mathbf{E}^{0} & = & 0,\\
\sum_{n=1}^{N}\left|k_{n}\right| & = & l.
\end{eqnarray*}
If the system does not have any resonance of order $l$ or lower, the Birkhoff normalization theorem \cite{classical3} guarantees that we can put $H$ in a normal form up to the remaining terms of order $l+1$
\begin{equation}
H^{*}=\sum_{n=1}^{N}E_{n}I_{n}+\sum_{n=1}^{l}H_{n}^{*}(I)+\mathcal{\mathcal{O}}(l+1).\label{HOK}
\end{equation}
In the case considered in this work, i.e., the case of a Hamiltonian function that comes from the expectation value of an operator, we
will see that there is no need to pay attention to possible resonances
in the energies due to the restricted form of the perturbation potentials.

We will now translate the Van Vleck-Primas quantum algorithm into
the classical Hamiltonian formalism. Let us note that the expectation
value of Eq.(\ref{DWH}) can be written as

\begin{eqnarray}
H^{*} & = & \sum_{n=0}^{\infty}\frac{1}{n!}\mathfrak{D}_{W}^{n}\left(H\right),\label{H**}
\end{eqnarray}
where $H^{*}=\left\langle \psi\right|\hat{\widetilde{H}}\left|\psi\right\rangle $,
$\mathfrak{D}_{W}=\left\{ \cdot,W^{*}\right\} $, and the generator
of the canonical transformation is $W^{*}=\left\langle \psi\right|\hat{W}\left|\psi\right\rangle $.
Both sides of Eq.(\ref{H**}) can be expanded in power series to get
a perturbation procedure. Alternatively, we can take the expectation
value of Eqs.(\ref{Qhomological}) to get the set of equations

\begin{equation}
\left\{ W_{n}^{*},H_{0}\right\} =\mathit{\Psi}_{n}-\left\langle \mathit{\Psi}_{n}\right\rangle ,\label{homological}
\end{equation}
where
\begin{eqnarray}
\mathit{\Psi}_{n} & = & \left\langle \psi\right|\hat{\Psi}_{n}\left|\psi\right\rangle ,\\
\left\langle \mathit{\Psi}_{n}\right\rangle  & = & \left\langle \psi\right|\pi\left(\hat{\Psi}_{n}\right)\left|\psi\right\rangle .
\end{eqnarray}
The Eqs.(\ref{homological}) are known in the classical context as
the homological equations \cite{classical3}. The functions $\mathit{\Psi}_{k}$ obtained as expectation values coincide with the ones found using
Hori's classical procedure \cite{classical0,classical3}. The first
three $\mathit{\Psi}$ functions are

\begin{eqnarray}
\mathit{\Psi}_{1} & = & V_{1},\nonumber \\
\mathit{\Psi}_{2} & = & V_{2}+\left\{ V_{1},W_{1}^{*}\right\} +\frac{1}{2}\left\{ \left\{ H_{0},W_{1}^{*}\right\} ,W_{1}^{*}\right\} ,\nonumber \\
\mathit{\Psi}_{3} & = & V_{3}+\left\{ V_{2},W_{1}^{*}\right\} +\left\{ V_{1},W_{2}^{*}\right\} +\frac{1}{2}\left\{ \left\{ V_{1},W_{1}^{*}\right\} W_{1}^{*}\right\} \nonumber \\
 &  & +\frac{1}{2}\left\{ \left\{ H_{0},W_{1}^{*}\right\} W_{2}^{*}\right\} +\frac{1}{2}\left\{ \left\{ H_{0},W_{2}^{*}\right\} W_{1}^{*}\right\} \nonumber \\
 &  & +\frac{1}{6}\left\{ \left\{ \left\{ H_{0},W_{1}^{*}\right\} ,W_{1}^{*}\right\} ,W_{1}^{*}\right\} .\label{PSI}
\end{eqnarray}
We point out that in Eqs.(\ref{homological}) and (\ref{PSI}) the
functions $H_{0}$, and $V_{n}$ are to be read as functions of $\left(\theta,I\right)$
in the same way as they originally depended on $(\theta^{(0)},I^{(0)})$. 

Let us note that, using Eq.(\ref{matrix}), the functions $\mathit{\Psi}_{n}$ can be written as 
\begin{equation}
\mathit{\Psi}_{n}(\theta,I)=\sum_{n,n'}\sqrt{I_{n}I_{n'}}\left\langle n'\right|\hat{\Psi}_{n}\left|n\right\rangle e^{i(\theta_{n'}-\theta_{n})}.\label{psinI}
\end{equation}
It remains for us to find an expression for $\left\langle \mathit{\Psi}_{n}\right\rangle .$
Using the definition (\ref{qaverage}), we can write

\begin{eqnarray}
\left\langle \mathit{\Psi}_{n}\right\rangle  & = & \lim_{\tau\rightarrow\infty}\frac{1}{\tau}\int_{0}^{\infty}\mathit{\Psi}_{n}(\theta_{k}-tE_{k},I_{k})\,dt\nonumber \\
 & = & \lim_{\tau\rightarrow\infty}\frac{1}{\tau}\int_{0}^{\infty}\left\langle \psi\right|e^{-i\hat{H}_{0}t}\hat{\Psi}_{n}e^{i\hat{H}_{0}t}\left|\psi\right\rangle dt\nonumber \\
 & = & \sum_{n,n'}\sqrt{I_{n}I_{n'}}\left\langle n'\right|\hat{\Psi}_{n}\left|n\right\rangle .\label{timeaverage}
\end{eqnarray}
The time average (\ref{timeaverage}) can be into a space average over a torus without changing the value
of $\left\langle \mathit{\Psi}_{n}\right\rangle $, though generally a non-resonance condition is required to for this to be true (page 286 of \cite{arnold}). As
a space average, $\left\langle \mathit{\Psi}_{n}\right\rangle $ reads

\begin{equation}
\left\langle \mathit{\Psi}_{n}\right\rangle =\frac{1}{\left(2\pi\right)^{N}}\int_{0}^{2\pi}\cdots\int_{0}^{2\pi}\mathit{\Psi}_{n}(\theta,I)\,d\theta_{1}\cdots d\theta_{N}.\label{spaceaverage}
\end{equation}
.

\[
\]

Now, let us note that, as a function of the new variables $\left(\theta,I\right)$,
the unperturbed Hamiltonian reads $H_{0}=\sum_{n=1}^{N}E_{n}^{0}I_{n}$.
This mean that Eq.(\ref{homological}) can be simplified to be

\begin{equation}
\sum_{n=1}^{N}E_{n}^{0}\frac{\partial W_{n}^{*}}{\partial\theta_{\mu}}=\mathit{\Psi}_{n}-\left\langle \mathit{\Psi}_{n}\right\rangle .\label{homological2}
\end{equation}
The Eq.(\ref{homological2}) can be solved by taking the Fourier expansion\footnote{Notice that the effect of $\left\langle \mathit{\Psi}_{n}\right\rangle $
is to eliminate the secular terms from $\mathit{\Psi}_{n}$ .} of $\mathit{\Psi}_{n}-\left\langle \mathit{\Psi}_{n}\right\rangle =\sum_{\mathbf{k}}A_{n,\mathbf{k}}e^{i\mathbf{k}\cdot\vec{\theta}}$.
The solution is \cite{classical3}

\begin{equation}
W_{n}^{*}=-\sum_{\mathbf{k}}\frac{A_{n,\mathbf{k}}e^{i\mathbf{k}\cdot\vec{\theta}}}{\mathbf{k}\cdot\mathbf{E}^{0}}.\label{W*}
\end{equation}
We can see that Eq.(\ref{W*}) has a small divisor problem if the
energies are resonant for some $\mathbf{k}$ unless $A_{n,\mathbf{k}}$
vanishes for such $\mathbf{k}$. There will be no diverging terms
in Eq.(\ref{W*}) in the non-degenerate case at hand. Indeed, we can
see from Eqs.(\ref{psinI}) and (\ref{timeaverage}) that 
\begin{equation}
W_{n}^{*}=-\sum_{n\neq n'}\frac{\sqrt{I_{n}I_{n'}}\left\langle n'\right|\hat{\Psi}_{n}\left|n\right\rangle }{E_{n'}^{0}-E_{n}^{0}}e^{i(\theta_{n'}-\theta_{n})}.\label{w*2}
\end{equation}

The Eq.(\ref{w*2}) completes our recasting of VVP perturbation theory
into a classical formalism. As there is a one-to-one correspondence
between a Hermitian operator and the real-valued function defined
by Eq.(\ref{g}), the quantum problem will be solved if the classical
generator functions are known. Given $W_{n}^{*}$, the matrix elements
of $\hat{W}_{n}$ can be read using the relation (\ref{matrix}).

Let us mention that if the energies do not have a resonance
of any order, the integrability of Hamiltonian (\ref{classicalh})
implies the convergence of classical perturbation series (and therefore,
the original quantum one) \cite{convergence}.

\[
\]
\section{Example: spin in a magnetic field}

In this section, we use an example to show how the generators of classical Lie transformations can be used to calculate the Berry phase of a quantum system. The procedure is based on two things: first, the VVP perturbation theory is related to the quantum geometric phase, see \cite{mio1}. Second, the generating functions of classical perturbation theory are related to the classical geometric phase (the Hannay angle) \cite{mio2}.  Hence, using the equivalence we have shown between the VVP theory and classical perturbation theory, we can use the latter to calculate the Berry phase. We mention that a classical approach to quantum geometric phases was discussed in \cite{Liu,Liu2}, however, our procedure is different.

In this section, the symbols $\theta$ and $\phi$ denote the polar
and azimuth angles of spherical coordinates. We use $\Theta$ for
the canonical angle variables. 

Consider the Hamiltonian of a single spin 1/2 interacting with a magnetic
field

\begin{equation}
\hat{H}=\frac{\mu}{2}\hat{\sigma}\cdot\mathbf{B}=\frac{\mu}{2}B\left(\begin{array}{cc}
\cos\theta & \sin\theta e^{-i\phi}\\
\sin\theta e^{i\phi} & -\cos\theta
\end{array}\right),\label{Hspin}
\end{equation}
where $\hat{\sigma}$ are the Pauli matrices. We then use the vector 

\begin{equation}
\left|\psi\right\rangle =\frac{q_{1}+ip_{1}}{\sqrt{2}}\left(\begin{array}{c}
0\\
1
\end{array}\right)+\frac{q_{2}+ip_{2}}{\sqrt{2}}\left(\begin{array}{c}
1\\
0
\end{array}\right)\label{psi1}
\end{equation}
to obtain the associated classical Hamiltonian function 

\begin{align}
\mathcal{H}=\left\langle \psi\right|\hat{H}\left|\psi\right\rangle = & \frac{\omega}{2}(q_{2}^{2}+p_{2}^{2}-q_{1}^{2}-p_{1}^{2})\cos\theta+\omega(q_{1}q_{2}+p_{1}p_{2})\sin\theta\cos\phi\nonumber \\
 & +\omega(p_{1}q_{2}-p_{2}q_{1})\sin\theta\sin\phi,
\end{align}
where

\begin{equation}
\omega=\frac{1}{2}\mu B.
\end{equation}

The transformation to angle action variable can be easily obtained
using the well-known eigenvectors of (\ref{Hspin}),

\begin{align}
\left|+\right\rangle  & =\left(\begin{array}{c}
\cos\frac{\theta}{2}\\
\sin\frac{\theta}{2}e^{i\phi}
\end{array}\right),\nonumber \\
\left|-\right\rangle  & =\left(\begin{array}{c}
-\sin\frac{\theta}{2}e^{-i\phi}\\
\cos\frac{\theta}{2}
\end{array}\right).\label{eigenV}
\end{align}
This allow us to rewrite $\left|\psi\right\rangle $ as 

\begin{equation}
\left|\psi\right\rangle =\frac{Q_{1}+iP_{1}}{\sqrt{2}}\left|-\right\rangle +\frac{Q_{2}+iP_{2}}{\sqrt{2}}\left|+\right\rangle .\label{psi2}
\end{equation}
Comparing the real and imaginary parts of Eqs (\ref{psi1}) and (\ref{psi2}),
we obtain 
\begin{align}
q_{1} & =Q_{1}\cos\frac{\theta}{2}+Q_{2}\sin\frac{\theta}{2}\cos\phi-P_{2}\sin\frac{\theta}{2}\sin\phi,\nonumber \\
p_{1} & =P_{1}\cos\frac{\theta}{2}+P_{2}\sin\frac{\theta}{2}\cos\phi+Q_{2}\sin\frac{\theta}{2}\sin\phi,\nonumber \\
q_{2} & =-Q_{1}\sin\frac{\theta}{2}\cos\phi+Q_{2}\cos\frac{\theta}{2}-P_{1}\sin\frac{\theta}{2}\sin\phi,\nonumber \\
p_{2} & =-P_{1}\sin\frac{\theta}{2}\cos\phi+P_{2}\cos\frac{\theta}{2}+Q_{1}\sin\frac{\theta}{2}\sin\phi.
\end{align}
Finally, the transformation 

\begin{align}
Q_{1} & =\sqrt{2I_{1}}\cos\Theta_{1},\;\;P_{1}=-\sqrt{2I_{1}}\sin\Theta_{1},\nonumber \\
Q_{2} & =\sqrt{2I_{2}}\cos\Theta_{2},\;\;P_{2}=-\sqrt{2I_{2}}\sin\Theta_{2}.
\end{align}
leads to

\begin{equation}
\mathcal{H}=\frac{\omega}{2}(I_{2}-I_{1}).
\end{equation}

Now, a small change in the orientation of the magnetic field makes the Hamiltonian vary, to first order, as

\begin{equation}
\mathcal{H}\rightarrow\mathcal{H}+\frac{\partial\mathcal{H}}{\partial\theta}\delta\theta+\frac{\partial\mathcal{H}}{\partial\phi}\delta\phi.
\end{equation}
We can then interpret the terms $\frac{\partial\mathcal{H}}{\partial\theta}\delta\theta$
and $\frac{\partial\mathcal{H}}{\partial\phi}\delta\phi$ as a small perturbation and apply classical perturbation theory to find the new
angle-action variables, hence the new eigenvectors. This may seem
unnecessary since (\ref{eigenV}) already gives the eigenvectors for
all angles. However, as mentioned, the generating functions contain information about the Berry phase. 

Now, we can read the generating functions associated with the variation
of each spherical angle directly from the homological equations. For example, the first-order homological equation for $\delta\theta$ is 
\begin{align}
\frac{\omega}{2}\frac{\partial W_{\theta}}{\partial\Theta_{2}}-\frac{\omega}{2}\frac{\partial W_{\theta}}{\partial\Theta_{1}} & =\frac{\partial\mathcal{H}}{\partial\theta}-\left\langle \frac{\partial\mathcal{H}}{\partial\theta}\right\rangle =-\frac{\omega}{2}\sqrt{I_{1}I_{2}}\sin\theta\cos(\Theta_{1}-\Theta_{2}+\phi),\label{homologicalw}
\end{align}
hence, we can write
\begin{align}
W_{\theta} & =\sqrt{I_{1}I_{2}}\sin\theta\sin(\Theta_{1}-\Theta_{2}+\phi). \label{generatingw}
\end{align}
A similar procedure gives
\begin{align}
W_{\phi} & =\sqrt{I_{1}I_{2}}\cos(\Theta_{1}-\Theta_{2}+\phi).\label{generatingw2}
\end{align}

Following \cite{mio2}, once we have the generating functions, the next step is the computation of the average of their Poisson bracket. The result is
\begin{equation}
\bigl\langle\bigl\{ W_{\theta},W_{\phi}\bigr\}\bigr\rangle=\frac{1}{2}I_{2}\sin\theta-\frac{1}{2}I_{1}\sin\theta.\label{curvature}
\end{equation}
Finally, taking the derivatives of (\ref{curvature}) by the corresponding action leads to the Berry curvature for the eigenvector

\begin{align*}
F_{1}^{Berry} & =-\frac{\partial}{\partial I_{1}}\bigl\langle\bigl\{ W_{\theta},W_{\phi}\bigr\}\bigr\rangle=\frac{1}{2}\sin\theta.\\
F_{2}^{Berry} & =-\frac{1}{2}\sin\theta.
\end{align*}

\section{Conclusions}

We have shown that the Van Vleck-Primas quantum perturbation theory
can be recast into the formalism of a classical Lie series as long as the Hamiltonian operator has a finite, discrete, and non-degenerate spectrum. On the other hand, complete non-resonance in the energies guarantees the convergence of the classical
(and therefore, the quantum) perturbation series.

Dealing only with finite-dimensional Hilbert spaces is extremely restrictive. Most applications to quantum systems require working with infinite-dimensional spaces. While there are a few works in the perturbation theory of classical systems with infinite but discrete degrees of freedom \cite{infinite1,infinite2}, this case is much less studied than its finite-dimensional counterpart. An extension to infinite-dimensional systems likely requires facing complications that do not appear in the finite-dimensional case. We leave such an attempt for future work. 

It is not a priori evident what advantages would come from giving a classical
perturbation treatment to a given quantum problem. Though this might
be due to our neglect of degeneracy and resonances (situations that are more common in the infinite-dimensional case). The classical perturbation
theory has a rich history of dealing with these issues, mainly due
to the particularities of the gravitational interaction in celestial
mechanics. For example, there are the Von Zeipel-Brouwer and Bohlin
theories (see \cite{classical3} for an exposition of both theories
and a large list of references). We can also mention the convergence
theorems of the Birkhoff normalization even for resonant cases \cite{convergence2,convergence3,convergence4}.

However,  our use of classical perturbation theory in quantum mechanics has already given a new approach to calculating the geometric phase. We hope that new procedures to deal with resonances in quantum systems can be developed by considering classical techniques.

\end{document}